\documentclass[showpacs,aps,preprint,prd]{revtex4}
\usepackage{graphicx}
\usepackage{amsmath}

\begin{document}
\title{\bf The Renormalization Group Limit Cycle for the $1/r^2$ Potential}

\author{Eric\ Braaten}

\affiliation{Department of Physics, 
	The Ohio State University, Columbus, OH 43210}

\author{Demian\ Phillips}

\affiliation{Department of Physics, 
	The Ohio State University, Columbus, OH 43210}

\date{\today}

\begin{abstract}
Previous work has shown that if an attractive $1/r^2$ potential 
is regularized at short distances 
by a spherical square-well potential, renormalization allows 
multiple solutions for the depth of the square well.  
The depth can be chosen to be a continuous
function of the short-distance cutoff $R$, 
but it can also be a log-periodic function of $R$
with finite discontinuities, corresponding to a renormalization 
group (RG) limit cycle.  We consider the regularization with a 
delta-shell potential.  In this case, the coupling constant is 
uniquely determined to be a log-periodic function of $R$ with 
infinite discontinuities, and an RG limit cycle is unavoidable.  
In general, a regularization with an RG limit cycle is selected 
as the correct renormalization of the $1/r^2$ potential by the 
conditions that the cutoff radius $R$ can be made arbitrarily small 
and that physical observables are reproduced accurately at all 
energies much less than $\hbar^2/mR^2$.
\end{abstract}

\pacs{11.10.Gh,03.65.-w,05.10.Cc}

%\vskip20pt

\maketitle

\section{Introduction}

The development of the {\it renormalization group} (RG) has had a profound 
impact on several subfields of physics \cite{Wilson:dy}.  
Many applications of the RG involve renormalization group flow 
towards a {\it fixed point} that is invariant under renormalization.
An example is critical phenomena in condensed matter physics, which 
can be understood in terms of renormalization group flow to a fixed point
in the infrared limit.  Another example is quantum chromodynamics (QCD), 
the quantum field theory that describes the strong interactions 
of elementary particles, which flows under 
renormalization to a fixed point in the ultraviolet limit.
A fixed point is the simplest topological feature that can be exhibited 
by an RG flow.  As pointed out by Wilson in 1970,
one of the next simplest possibilities is a {\it limit cycle},
a closed curve that is invariant under renormalization \cite{Wilson:1970ag}.  
The limit cycle is characterized by a {\it discrete scaling symmetry}: 
the renormalization group flow executes a complete cycle
around the curve every time the cutoff changes by a 
multiplicative factor $\lambda$ called the {\it discrete scaling factor}.  
The discrete scaling symmetry is reflected in {\it log-periodic} 
behavior of physical observables as functions of the momentum scale.
The possibility of RG limit cycles has received little attention 
until recently, partly because of the scarcity of compelling examples.
One physical example that was identified long ago is the problem 
of identical bosons with large scattering length $a$ \cite{Albe-81}.  
In the limit $a \to \pm \infty$, there is an accumulation of 
3-body bound states near threshold with binding energies differing by
multiplicative factors of $\lambda^2 \simeq 515.03$ \cite{Efimov70}.
This phenomenon, which is called the Efimov effect, can be understood 
in terms of a renormalization group limit cycle with 
discrete scaling factor $\lambda \simeq 22.7$.
This application has been made more compelling by Bedaque, 
Hammer, and van Kolck, who reformulated the problem using 
effective field theory \cite{Bedaque:1998kg}. 
Other examples of renormalization group limit cycles have recently begun to
emerge.  There are discrete Hamiltonian systems 
that exhibit RG limit cycles in appropriate continuum limits 
\cite{Glazek:2002hq,LeClair:2002ux}.
LeClair, Roman, and Sierra have identified a two-dimensional field theory 
whose renormalization involves an RG limit cycle \cite{Leclair:2003xj} 
in apparent contradiction to Zamolodchikov's C theorem 
\cite{Zamolodchikov:gt}.  It has even been conjectured that 
QCD has an infrared RG limit cycle 
at special values of the quark masses \cite{Braaten:2003dw}.

These examples suggest that RG limit cycles may play a more important
role in physics than previously realized.  They provide motivation 
for studying simple examples of RG limit cycles.  
The simplest example is the quantum mechanics of a particle in a
potential whose long-range behavior is $1/r^2$.
This problem has been studied previously within the renormalization group
framework by two different groups using a spherical square-well 
regularization potential \cite{Beane:2000wh,Bawin:2003dm}.
Beane et al.~\cite{Beane:2000wh} showed that there are infinitely many 
choices for the coupling constant of the square-well 
potential, including a continuous function of the short-distance 
cutoff $R$ and a log-periodic function of $R$ 
with discontinuities which corresponds to an RG limit cycle.  
Bawin and Coon \cite{Bawin:2003dm} presented a closed-form solution 
for the coupling constant that is log-periodic, which suggests that the 
choice with the RG limit cycle is in some sense natural.  

In this paper, we clarify the role of RG limit cycles in the 
renormalization of the $1/r^2$ potential.  We begin in Section 
\ref{sec:renorm} by summarizing how renormalization theory can 
be applied to the $1/r^2$ potential.  In Section \ref{sec:sqwreg}, 
we reconsider the spherical square-well regularization potential and 
calculate the bound-state spectrum for alternative choices of the 
coupling constant.  In Section \ref{sec:dshreg}, we consider a 
spherical delta-shell regularization potential.  In this case, 
the coupling constant is uniquely determined and is governed by 
an RG limit cycle. We discuss our results in Section \ref{sec:disc} 
and identify the criterion that selects the regularization 
with the RG limit cycle as the correct renormalization of 
the $1/r^2$ potential.

\section{Renormalization of the $1/r^2$ Potential}
\label{sec:renorm}

We consider a particle in a spherically-symmetric potential $V(r)$ 
that is attractive and proportional to $1/r^2$ 
for $r$ greater than some radius $R_{\rm min}$:
\begin{align}
V(r) & = - \left( \mbox{$1\over 4$} + \nu^2 \right) {\hbar^2 \over 2 m r^2}
	&& r > R_{\rm min},
\nonumber
\\
     & = V_{\rm short}(r) 
	&& r \le R_{\rm min},
\label{Vtrue}
\end{align}
where $\nu$ is a positive parameter.  The coefficient of the 
short-distance potential is written as ${1\over 4} + \nu^2$ 
because $\nu^2 = 0$ is the critical value above which the potential 
is too singular for the problem to be well-behaved in the limit 
$R_{\rm min} \rightarrow 0$.
We will not specify the short-distance potential $V_{\rm short}(r)$.
The potential $V(r)$ has infinitely-many arbitrarily-shallow S-wave 
bound states 
with an accumulation point at the scattering threshold $E=0$.  
As the threshold is approached, the ratio of the binding energies 
of successive states approaches $\lambda^2 = e^{2\pi/\nu}$.  
The asymptotic spectrum near the threshold therefore has the form
\begin{equation}
E_n \longrightarrow - {\hbar^2 \kappa_*^2 \over m} \left(e^{-2\pi/\nu}\right)^{n-n_*},
\label{E-low}
\end{equation}
where  $n_*$ is an integer that can be chosen for convenience
and $\kappa_*$ is determined up to a multiplicative factor of 
$e^{\pi/\nu}$ by the short-distance potential.
This geometric spectrum reflects an asymptotic discrete scaling symmetry
in which the distance from the origin is rescaled by the discrete 
scaling factor 
$\lambda = e^{\pi/\nu}$.  One might have expected an approximate 
continuous scaling symmetry because the long-distance potential is
scale-invariant, but the continuous scaling symmetry is broken 
to a discrete subgroup by the boundary conditions provided by the 
short-distance potential.  This is an example of a quantum mechanical 
{\it anomaly}.  

Although renormalization theory was originally introduced to attack 
very different problems \cite{Wilson:dy}, it can also be applied to 
nonrelativistic quantum mechanics \cite{Adhikari:1997dz}.  
A particularly convenient way of implementing renormalization theory 
in quantum mechanics is within an effective theory framework, 
which allows a systematically improvable description of the system 
at low energies $E$ satisfying $|E| \ll \hbar^2 / m R_{\rm min}^2$ 
\cite{Lepage:1997cs}.
Renormalization can be implemented in this problem by introducing 
a cutoff radius $R$ satisfying $R > R_{\rm min}$ 
and replacing the potential in the region $0 < r < R$ by a 
regularization potential $V_{\rm reg}(r;\lambda)$
that depends on a tuning parameter $\lambda$:
\begin{align}
V(r) & = - \left( \mbox{$1\over 4$} + \nu^2 \right) {\hbar^2 \over 2 m r^2}
	&& r > R,
\nonumber
\\
     & = V_{\rm reg}(r;\lambda(R))
	&& r \le R.
\label{Vreg}
\end{align}
Some quantity involving low energies $|E| \ll \hbar^2/mR^2$,
such as the energy eigenvalue of a very shallow bound state,
is selected as a matching variable.
The parameter $\lambda(R)$ in the regularization potential is then tuned 
so that the value of the matching variable in the true theory 
with potential (\ref{Vtrue}) is reproduced by the theory with the 
regularized potential (\ref{Vreg}).  Renormalization theory guarantees 
that other low-energy observables involving energies 
satisfying $|E| \ll \hbar^2/mR^2$
will also be reproduced correctly by the regularized theory 
up to corrections of order $E mR^2/\hbar^2$.
As $R$ is decreased, the errors decrease as $R^2$
until $R$ reaches $R_{\rm min}$.
If $R$ is decreased below $R_{\rm min}$, 
there is no further decrease in the errors.

A particularly convenient choice for the matching quantity 
is the zero-energy wavefunction.  The stationary 
Schroedinger equation for a radial wavefunction $u(r)/r$ 
with zero angular momentum and energy eigenvalue $E$ is 
\begin{equation}
\left(-\frac{\hbar^2}{2m}\frac{d^2}{dr^2}+V(r)\right) u(r) = E u(r).
\end{equation}
The $E=0$ solution in the $1/r^2$ region of the true potential $V(R)$
has the form
%%%%%%%%%%%%%%%%%%%%%%%%%%%%%%%%%%%%%%%%%%%%%%%%%%%%%%%%%%%%%%%%%%%%%%%%%%%%%%%%%%%%%%%%%%%%%%%%%%
\begin{figure}[htbp]
\begin{center}
$ \begin{array}{cc}
\includegraphics[width=2.5in]{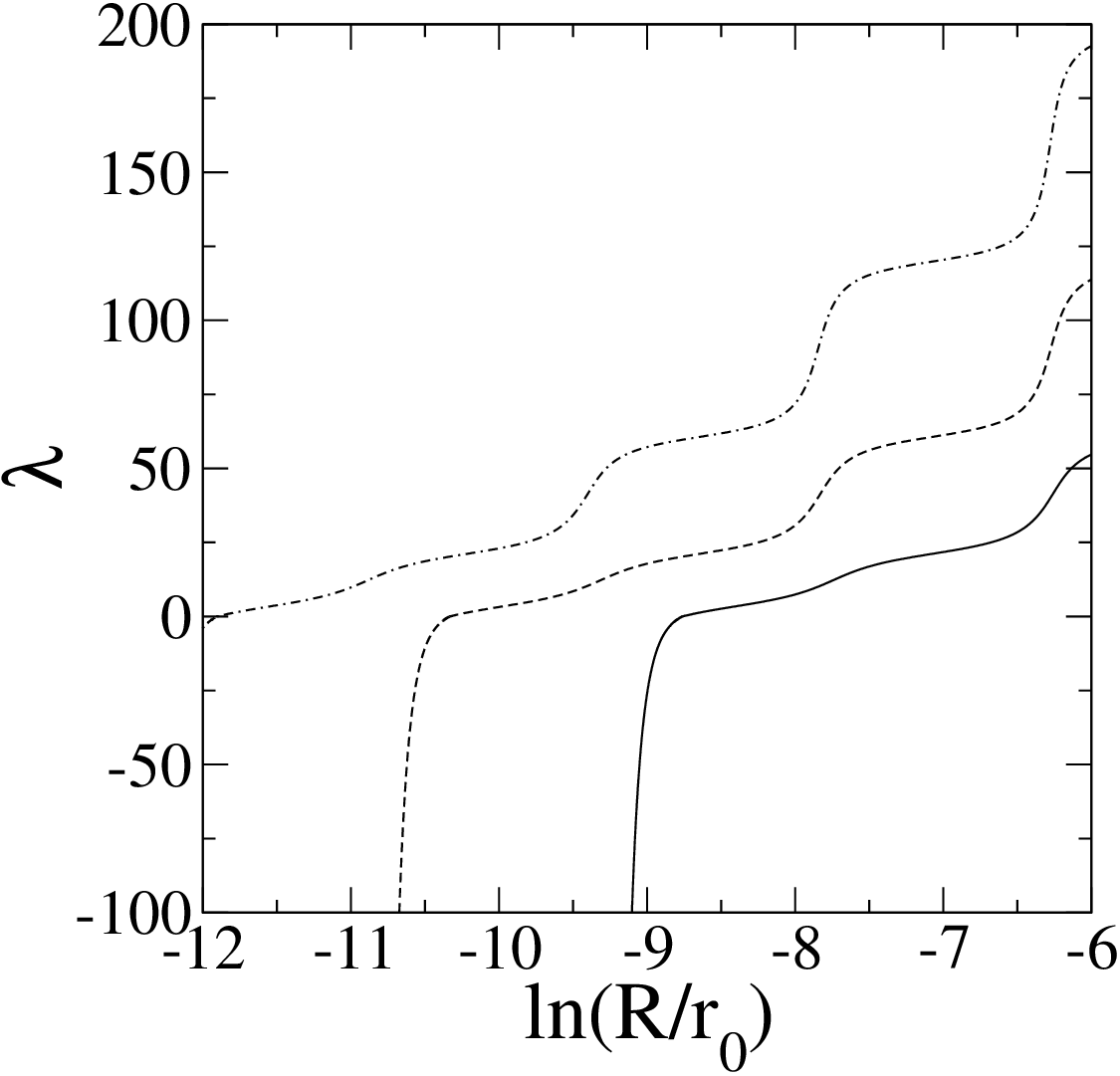}
\hspace{.5in} &
\includegraphics[width=2.5in]{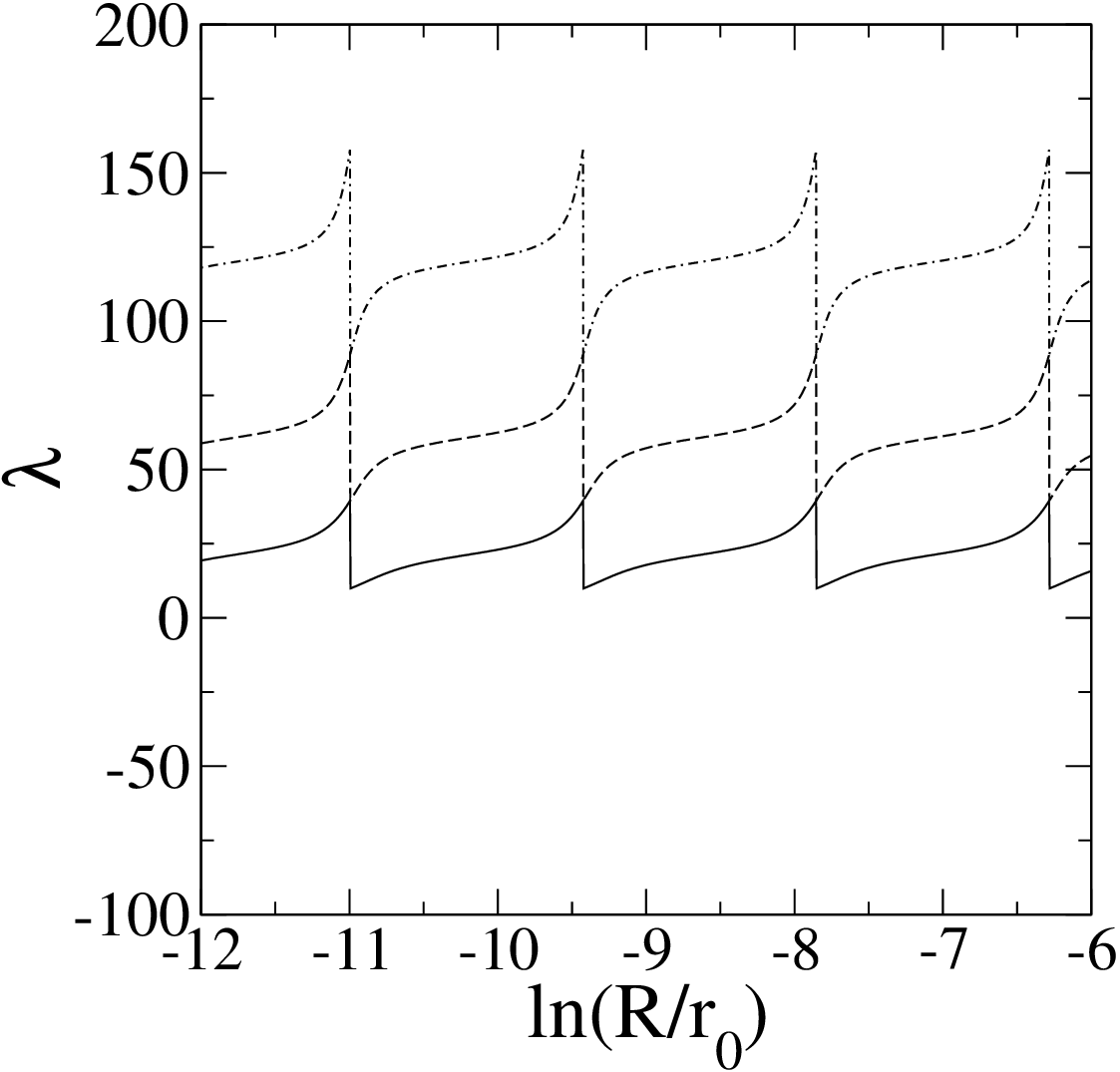} \\
{\mathrm (a)} \hspace{.5in} & {\mathrm (b)} 
\end{array} $
\end{center}
\caption{Several branches of the coupling constant $\lambda(R)$ 
of the square-well regularization potential for $\nu=2$ as a 
function of $\ln(R/r_0)$ for $(a)$ continuous $\lambda(R)$ 
and $(b)$ log-periodic $\lambda(R)$.} 
\label{fig:lambda-sqw}
\end{figure}
%%%%%%%%%%%%%%%%%%%%%%%%%%%%%%%%%%%%%%%%%%%%%%%%%%%%%%%%%%%%%%%%%%%%%%%%%%%%%%%%%%%%%%%%%%%%%%%%%%
%
\begin{equation}
u(r) = A r^{1/2} \sin \left[ \nu \ln (r/r_0) \right] 
\hspace{1cm} r > R_{\rm min}.
\label{u0}
\end{equation}
The parameter $r_0$ is the position of one of the nodes of the wavefunction.  
It is determined up to a multiplicative factor $e^{\pi/\nu}$ by the 
short-distance potential $V_{\rm short}(r)$. 

\section{Square-Well Regularization}
\label{sec:sqwreg}

In two previous studies of the $1/r^2$ potential using 
renormalization theory, the regularization potential 
was chosen to be a spherical square well \cite{Beane:2000wh,Bawin:2003dm}:
\begin{equation}
V_{\rm reg}(r;\lambda) = - \lambda {\hbar^2 \over 2 m R^2}
\hspace{1cm} r < R.
\end{equation}
The ``coupling constant'' $\lambda$ is dimensionless.  
The condition that the regularized potential reproduce the 
zero-energy wavefunction (\ref{u0}) at distances $r > R$ is
\begin{equation}
\lambda^{1/2} \cot(\lambda^{1/2}) = 
\mbox{$1\over2$} +\nu \cot \left[\nu \ln (R/r_0)\right] .
\label{lambda-sqw}
\end{equation}
This equation applies not only for $\lambda > 0$, but also for 
$\lambda < 0$, in which case
$\lambda^{1/2} \cot(\lambda^{1/2})=|\lambda|^{1/2} \coth(|\lambda|^{1/2})$.
For any value of $R$, the transcendental equation (\ref{lambda-sqw}) 
has infinitely many roots.

In Ref.~\cite{Beane:2000wh}, Beane et al.~pointed out that $\lambda(R)$
could be chosen to be continuous.  For the case $\nu=2$, 
three branches of continuous $\lambda(R)$ are illustrated in 
%%%%%%%%%%%%%%%%%%%%%%%%%%%%%%%%%%%%%%%%%%%%%%%%%%%%%%%%%%%%%%%%%%%%%%%%%%%%%%%%%%%%%%%%%%%%%%%%%%
\begin{figure}[htbp]
\begin{center}
$ \begin{array}{cc}
\includegraphics[width=2.5in]{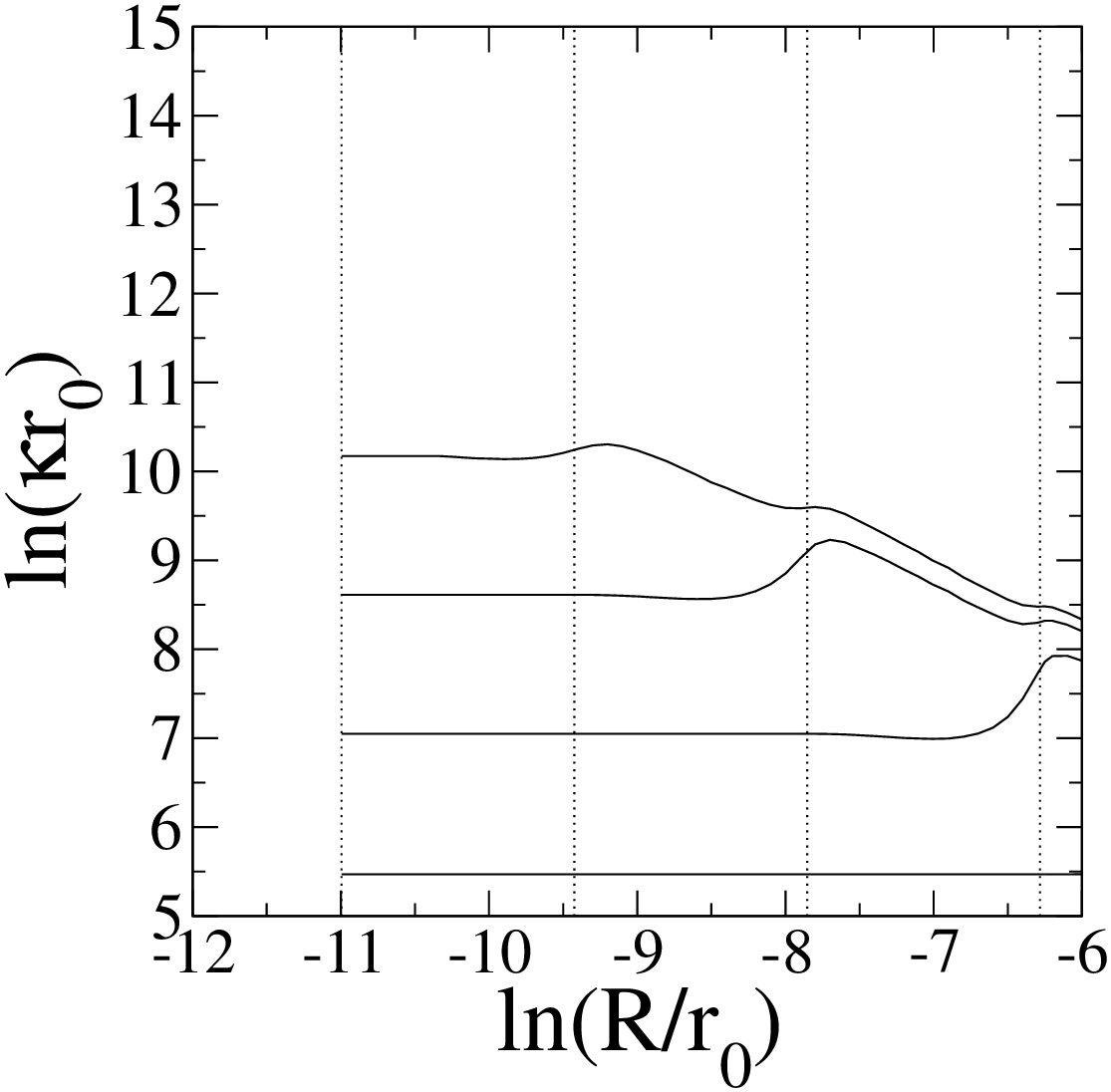}
\hspace{.5in} &
\includegraphics[width=2.5in]{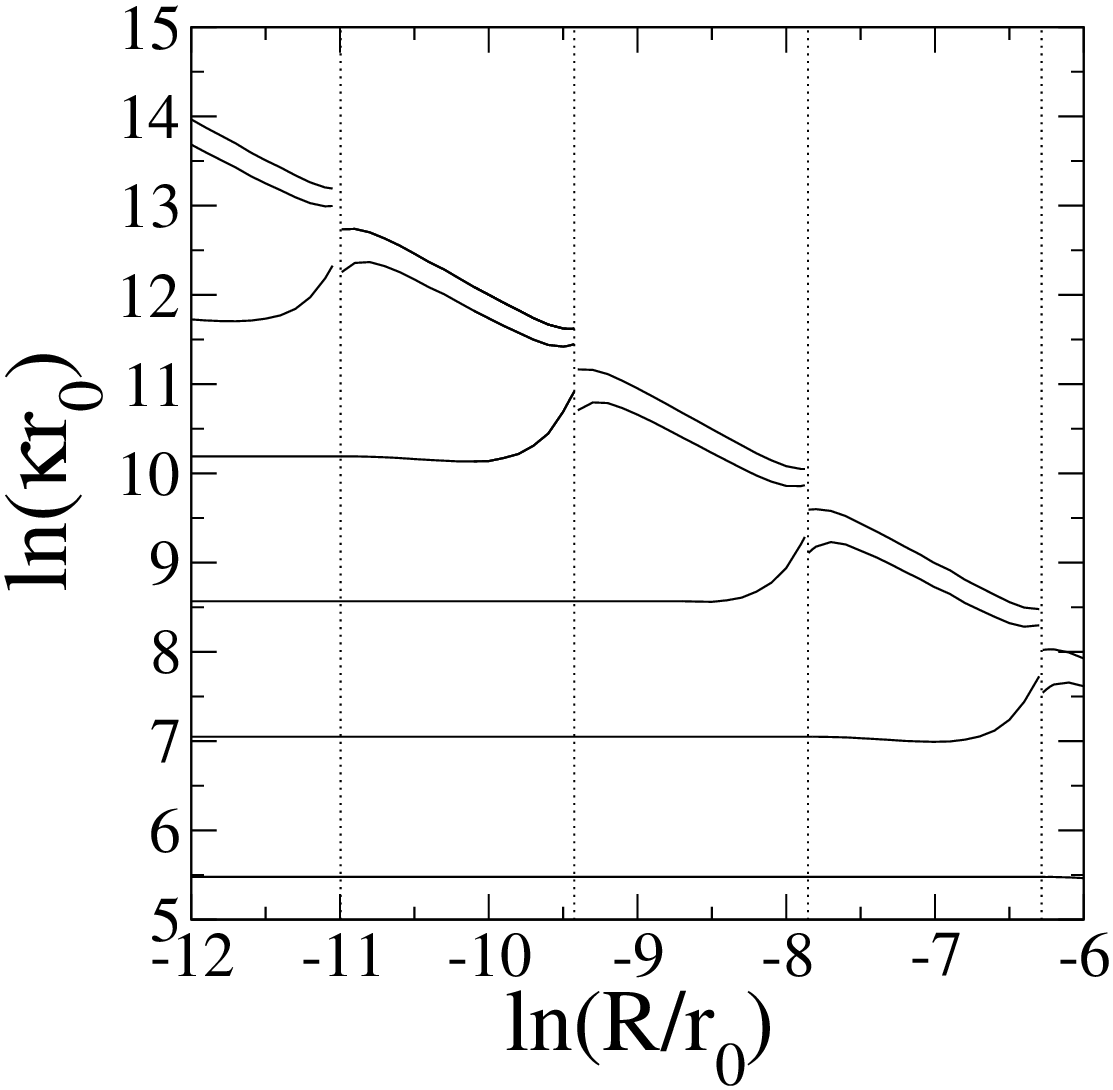} \\
{\mathrm (a)} \hspace{.5in} & {\mathrm (b)}
\end{array} $
\end{center}
\caption{The spectrum of the deepest bound states for $\nu=2$ 
and the square-well regularization potential as a function of 
ln$(R/r_0)$ for $(a)$ continuous $\lambda(R)$ given by the 
middle branch of Fig. 1$(a)$ and $(b)$ log-periodic $\lambda(R)$ 
given by the middle branch of Fig. 1$(b)$.}
\label{fig:spectrum-sqw}
\end{figure}
%%%%%%%%%%%%%%%%%%%%%%%%%%%%%%%%%%%%%%%%%%%%%%%%%%%%%%%%%%%%%%%%%%%%%%%%%%%%%%%%%%%%%%%%%%%%%%%%%%
Fig.~\ref{fig:lambda-sqw}(a).  The coupling 
constant $\lambda(R)$ decreases monotonically 
as the cutoff radius $R$ decreases.  Once it becomes negative, 
it decreases very rapidly and reaches $-\infty$ at a finite value 
of $R$ given by
\begin{equation}
R^{(M)} = \left(e^{-\pi/\nu}\right)^M r_0,
\label{infty-sqw}
\end{equation}
where $M$ is an integer.  Thus, if $\lambda(R)$ is continuous, 
there is a lower bound on the cutoff radius $R$.  
The three branches in Fig.~\ref{fig:lambda-sqw}(a) correspond to 
three consecutive values of $M$.  
The authors of Ref.~\cite{Beane:2000wh} also pointed out that $\lambda(R)$
could equally well be chosen to jump discontinuously between
the branches of the solutions to (\ref{lambda-sqw}) at arbitrary values 
of $R$ without affecting the observables at 
extremely low energies.  One particular choice would be to have 
$\lambda(R)$ jump up to the next branch every time $R$ decreases 
by a factor of $e^{\pi/\nu}$.  This choice corresponds to an RG limit cycle.

In Ref.~\cite{Bawin:2003dm}, Bawin and Coon presented  
a closed-form solution to (\ref{lambda-sqw}) 
that depends on an integer parameter $n$.
The resulting coupling constants $\lambda_n(R)$ are log-periodic functions 
of $R$.  For the case $\nu=2$, three branches of log-periodic 
$\lambda(R)$ are illustrated in Fig.~\ref{fig:lambda-sqw}(b).  
They have finite discontinuities at the specific values of $R$ 
given by (\ref{infty-sqw}), which differ by multiplicative factors 
of $e^{\pi/\nu}$. Such a choice of the solution to (\ref{lambda-sqw}) 
corresponds to a renormalization group limit cycle with 
discrete scaling factor $e^{\pi/\nu}$.

We now consider the bound-state spectrum.  The equation for the 
bound-state energy eigenvalues $E_n = -\hbar^2 \kappa_n^2 / 2m$ is
\begin{equation}
\frac{1}{2} + \kappa R \frac {K'_{i\nu} (\kappa R)}{K_{i\nu} (\kappa R)} 
= (\lambda(R) - \kappa^2 R^2)^{1/2} \cot \left( (\lambda(R) - \kappa^2 R^2)^{1/2} \right),
\label{spectrum-sqw}
\end{equation}
where $K_{i\nu}(z)$ is a modified Bessel function with an imaginary index.  
The spectrum of very shallow bound states is almost completely independent 
of $R$ and has the form (\ref{E-low}).  The parameter $\kappa_*$ 
in (\ref{E-low}) is related to the parameter $r_0$ in the
 zero-energy wavefunction by 
\begin{equation}
\kappa_* e^{n \pi/\nu} =  {2 \over r_0} e^{\arg\Gamma(1+i\nu)/\nu},
\label{kappa-r0}
\end{equation}
where $n$ is an integer that depends on the choices for $\kappa_*$ 
and $r_0$, both of which are defined only up to multiplicative 
factors of $e^{\pi/\nu}$.  The spectrum of deeper bound states 
depends on $R$ and on the choice of the branch for $\lambda(R)$.
The spectrum of the deepest bound states for $\nu=2$ and for 
continuous $\lambda(R)$ given by the middle branch in 
Fig.~\ref{fig:lambda-sqw}(a) is shown in Fig.~\ref{fig:spectrum-sqw}(a).
The curves cannot be extended below the value $R^{(M)}$ given by 
(\ref{infty-sqw}) because $\lambda(R)$
reaches $-\infty$ at that point.  The binding energies are all 
continuous functions of $R$.  For $R > R^{(M-1)}$, the order of 
magnitude of the deepest binding energy is $\hbar^2/mR^2$.  
The spectrum of the deepest bound states for $\nu=2$ and for 
log-periodic $\lambda(R)$ given by the middle branch in 
Fig.~\ref{fig:lambda-sqw}(b) is shown in Fig.~\ref{fig:spectrum-sqw}(b).  
At each of the values of $R$ at which $\lambda(R)$ jumps 
discontinuously, a new deepest bound state appears in the spectrum.
As $R$ decreases further, the third deepest bound state rapidly 
approaches its asymptotic value given by (\ref{E-low}) 
and (\ref{kappa-r0}).

\section{Delta-Shell Regularization}
\label{sec:dshreg}

\begin{figure}
\includegraphics[angle=0, width=6cm]{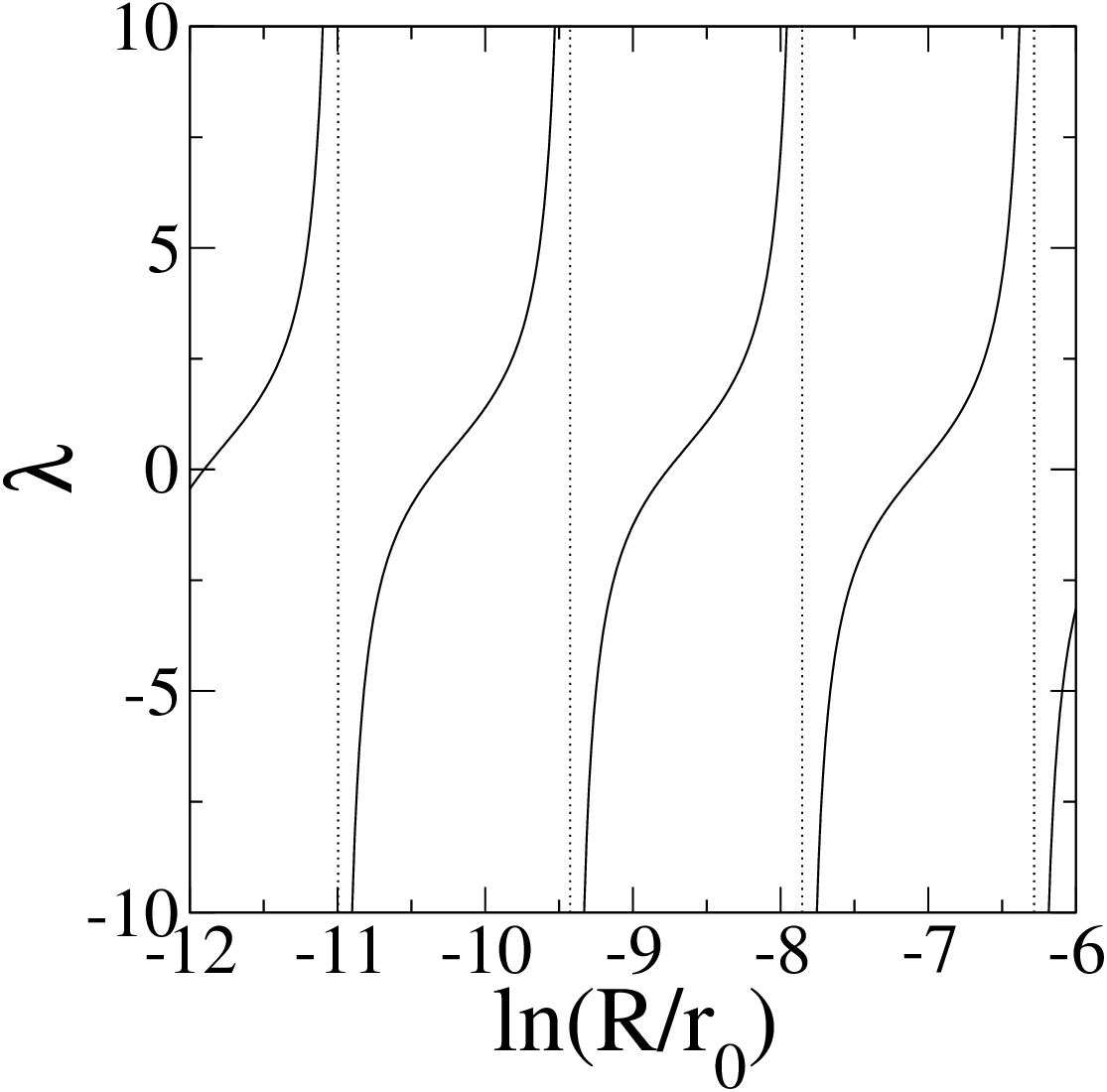}
\caption{The coupling constant $\lambda(R)$ for $\nu = 2$ and
the delta-shell regularization potential as a function of ln$(R/r_0)$ .}
\label{fig:lambda-dsh}
\end{figure}

Renormalization theory is designed to give results for low-energy 
observables that are independent of the regularization potential.  
One regularization potential that is particularly convenient is 
the spherical delta-shell consisting 
of a delta function concentrated on a shell with radius $r=R^-$ 
infinitesimally close to but smaller than $R$:
\begin{equation}
V_{\rm reg}(r;\lambda) = -\lambda \frac{\hbar^2}{2mR} \delta (r-R^-)
\hspace{1cm} r \le R.
\end{equation}
The coupling constant $\lambda$ is dimensionless.  The radial 
wavefunction $u(r)/r$ must be continuous at $r=R$.  
Another boundary condition at $r=R$ is obtained by integrating 
the Schroedinger equation over an infinitesimal region including $r = R$:
\begin{equation}
\lim_{r \to R^+} r {u'(r) \over u(r)} 
- \lim_{r \to R^-} r {u'(r) \over u(r)} = -\lambda(R).
\label{bc2}
\end{equation}
The scattering solution for energy $E=\hbar^2 k^2/2m$ has the form 
\begin{align}
u(r) & =  r^{1/2} \left[A_+ J_{i \nu}(kr) + A_- J_{-i \nu}(kr) \right]
	&& r > R ,
\nonumber
\\
     & =  A' \sin(kr) 
	&& r < R.
\end{align}
This reduces to the zero-energy solution (\ref{u0}) as 
$k \longrightarrow 0$ if the limiting behavior of the coefficients is 
$A_{\mp} \longrightarrow \mp \frac{1}{2} iA\left(2/kr_0 \right)^{\pm i \nu}$.  
Applying the boundary condition 
(\ref{bc2}) to this solution and taking the limit $k \longrightarrow 0$, we
determine the coupling constant $\lambda(R)$:
\begin{equation}
\lambda(R) = \frac{1}{2} -\nu \cot \left[\nu \ln (R/r_0) \right] .
\label{lam-delta}
\end{equation}
The coupling constant (\ref{lam-delta}) is a single-valued function of $R$, 
in contrast to the case of the square-well regularization where the
coupling constant has infinitely many branches.
As shown in Fig.~\ref{fig:lambda-dsh}, $\lambda(R)$ is a 
log-periodic function of $R$ with infinite discontinuities.
It jumps discontinuously 
from $+\infty$ to $-\infty$
as $R$ decreases through the critical values given by (\ref{infty-sqw}).

%%%%%%%%%%%%%%%%%%%%%%%%%%%%%%%%%%%%%%%%%%%%%%%%%%%%%%%%%%%%%%%%%%%%%%%%%%%%%%%%%%%%%%%%%%%%%%%%%%
%%%%%%%%%%%%%%%%%%%%%%%%%%%%%%%%%%%%%%%%%%%%%%%%%%%%%%%%%%%%%%%%%%%%%%%%%%%%%%%%%%%%%%%%%%%%%%%%%%

\begin{figure}
\includegraphics[angle=0, width=6cm]{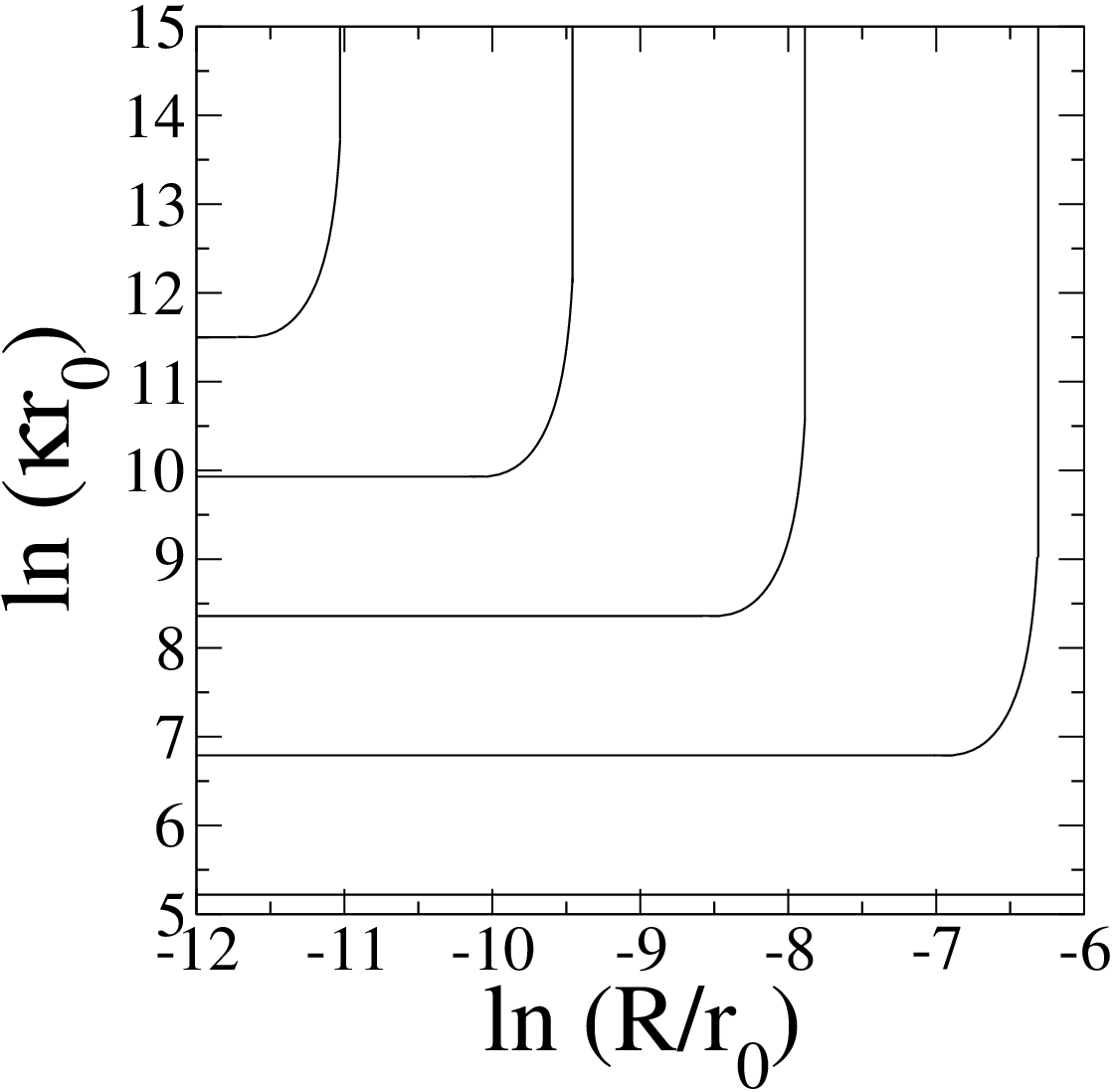}
\caption{The spectrum of the deepest bound states for $\nu = 2$ 
and the delta-shell regularization potential as a function 
of $\ln(R/r_0)$.}
\label{fig:spectrum-dsh}
\end{figure}

We now consider the bound state spectrum.  
The radial wavefunction for a negative energy 
$E= -\hbar^2 \kappa^2 / 2m$ has the form 
\begin{align}
u(r) & = B r^{1/2} K_{i\nu} (\kappa r) 
	&& r > R, 
\nonumber
\\
     & = B' \sinh(\kappa r) 
	&& r < R .
\end{align}
Using the boundary condition (\ref{bc2}),
we find that the equation for the bound-state energy eigenvalues 
$E_n=-\hbar^2\kappa_n^2/2m$ is
\begin{equation}
\frac{1}{2} + \kappa R \frac{K'_{i\nu} (\kappa R)}{K_{i\nu} (\kappa R)}
- \kappa R \coth (\kappa R) = -\lambda (R).
\label{bs-delta}
\end{equation}
The spectrum of very shallow bound states is almost completely 
independent of $R$ and has the form (\ref{E-low}) with $\kappa_*$ 
given by (\ref{kappa-r0}).  The spectrum for the deepest bound states 
is illustrated in 
Fig.~\ref{fig:spectrum-dsh}.  At the critical values of $R$ given by 
(\ref{infty-sqw}), where $\lambda(R)$ changes
discontinuously from $-\infty$ to $+\infty$, a new bound state 
with infinitely deep binding energy 
$\kappa = +\infty$ emerges.  As $R$ decreases further, 
that binding energy rapidly approaches its asymptotic value given 
by (\ref{E-low}) and (\ref{kappa-r0}).  Comparing with 
Fig.~\ref{fig:spectrum-sqw}(b), we see that the deepest bound state 
corresponds to the third deepest bound state for the 
square-well regularization with an RG limit cycle.

%%%%%%%%%%%%%%%%%%%%%%%%%%%%%%%%%%%%%%%%%%%%%%%%%%%%%%%%%%%%%%%%%%%%%%%%%%%%%%%%%%%%%%%%%%%%%%%%%%
%%%%%%%%%%%%%%%%%%%%%%%%%%%%%%%%%%%%%%%%%%%%%%%%%%%%%%%%%%%%%%%%%%%%%%%%%%%%%%%%%%%%%%%%%%%%%%%%%%

\section{Discussion}
\label{sec:disc}

We have studied the renormalization of an attractive $1/r^2$ potential 
using two regularization potentials:
a spherical square well as in Refs. \cite{Beane:2000wh} 
and \cite{Bawin:2003dm} and a spherical delta shell.  In the case
of the delta-shell potential, the coupling constant $\lambda$ is 
necessarily a log-periodic function 
of the cutoff radius $R$ with infinite discontinuities.  
It is governed by a renormalization group (RG) limit
cycle.  In the case of the square-well potential, there is much more 
freedom because there are infinitely many branches for the 
coupling constant $\lambda$.  It might seem natural to choose 
$\lambda(R)$ to be a continuous function of $R$, but this 
choice has some drawbacks.  Since $\lambda(R)$ diverges to $-\infty$ 
at a finite value of $R$, the cutoff radius cannot be
decreased below this value.  The choice of continuous $\lambda(R)$ 
also imposes an upper bound on the binding energy of 
the deepest bound state.  Alternatively, the coupling constant 
can be chosen to be a log-periodic function of $R$ with 
finite discontinuities, corresponding to an RG limit cycle.  
With this choice, the cutoff can be decreased to arbitrarily
short distances and there is no upper bound on the binding energies.  

If the value of the physical short-distance cutoff $R_{\rm min}$ is fixed 
and known in advance, the choice between a log-periodic $\lambda(R)$ 
with an RG limit cycle and continuous $\lambda(R)$ is only a matter of taste.  
For continuous $\lambda(R)$, one can simply choose a branch of the 
coupling constant such that the minimal value of the cutoff radius is 
smaller than $R_{\rm min}$.  However, if $R_{\rm min}$ is not known 
or if it can be varied, the continuous choice of 
$\lambda(R)$ will break down if $R_{\rm max}$ happens to be smaller 
than the minimal cutoff radius.  The choice of log-periodic $\lambda(R)$ 
with an RG limit cycle guarantees that $R$ can be made arbitrarily small 
and the effective potential still reproduces accurately all physics 
involving energies much smaller than $\hbar^2/mR^2$.  
This criterion selects the regularization with an RG limit cycle 
as the correct renormalization of the $1/r^2$ potential.  

The RG limit cycle for the $1/r^2$ potential has also been studied 
recently using flow equations for RG transformations \cite{Mueller:2004ejm}.

\acknowledgments

We thank K.G. Wilson for valuable discussions.  
This research was supported in part by the Department of Energy 
under grant DE-FG02-91-ER4069.

\end{document}